\RequirePackage{amsmath} 
\documentclass[12pt]{iopart}
\usepackage[utf8]{inputenc} 
\usepackage[T1]{fontenc} 
\usepackage{textcomp,lmodern} 
\usepackage[normalem]{ulem} 
\usepackage{amsthm, amssymb, amsmath} 
\usepackage{bbm} 
\DeclareMathOperator{\Spur}{Tr} %

\renewcommand{\Re}[0]{\ens{\mathrm{Re}}} 
\renewcommand{\Im}[0]{\ens{\mathrm{Im}}} 
\newcommand{\ens}[0]{\ensuremath} 
\newcommand{\anfEngl}[1]{``#1''} 
\newcommand{\ket}[1]{\ens{|#1\rangle}} 
\newcommand{\bra}[1]{\ens{\langle#1|}} 
\newcommand{\pr}[1]{\ens{\ket{#1}\bra{#1}}} 
\newcommand{\eqv}[0]{\ens{\Leftrightarrow}} 
\newcommand{\nach}[0]{\ens{\rightarrow}} 
\newcommand{\Mge}[2]{\ens{\left\lbrace #1|\,#2 \right\rbrace}} 
\newcommand{\Mg}[1]{\ens{\left\lbrace #1 \right\rbrace}} 
\newcommand{\norm}[1]{\ens{\left\lVert#1\right\rVert}} 
\newcommand{\betrag}[1]{\ens{|#1|}} 
\newcommand{\Fkt}[3]{\ens{#1 : #2 \nach #3}} 
\newcommand{\Eins}[0]{\ens{\mathbbm{1}}} 
\newcommand{\df}[1]{\mathrm{d}#1}
\renewcommand{\phi}[0]{\ens{\varphi}} 
\renewcommand{\theta}[0]{\ens{\vartheta}} 
\newcommand{\my}[0]{\ens{\mu}} 
\newcommand{\ksi}[0]{\ens{\xi}}
\newcommand{\cH}[0]{\ens{\mathcal{H}}}

\newcommand{\N}[0]{\ens{\mathbb{N}}} 

\newcommand{\R}[0]{\ens{\mathbb{R}}}
\newcommand{\C}[0]{\ens{\mathbb{C}}}
\newcommand{\iE}[0]{\ens{\mathrm{i}}} 
\newcommand{\EZ}[0]{\ens{\mathrm{e}}} 
\newcommand{\da}[0]{\ens{\,\df{{}^2\alpha}}}
\newcommand{\dxdp}[0]{\ens{\,\df{x}\,\df{p}}}
\newcommand{\D}[0]{\ens{\hat{D}}} 
\newcommand{\Sq}[0]{\ens{\hat{S}}}
\newcommand{\erz}[0]{\ens{\hat{a}^\dagger}}
\newcommand{\vnt}[0]{\ens{\hat{a}}}
\newcommand{\anz}[0]{\ens{\hat{n}}}

\newcommand{\PhV}[0]{\ens{W}} 
\newcommand{\pers}[1]{\textsc{#1}} 
\newcommand{\persbf}[1]{#1} 
\newcommand{\aut}[2]{\textsc{#1 \uline{#2}}} 
\usepackage{stringstrings}
\newcommand\removelowercase[2][v]{\Treatments{1}{0}{0}{0}{0}{1} \substring[#1]{#2}{1}{100} \defaultTreatments}
\renewcommand{\aut}[2]{\textsc{#2\removelowercase{#1}}\hspace{-0.1cm}} 
\newcommand{\autgr}[2]{\textsc{#1 and #2}} 
\newcommand{\tit}[1]{\textit{#1}} 
\newcommand{\zeit}[1]{#1} 
\newcommand{\Band}[1]{\textbf{#1}} 
\newcommand{\Heft}[1]{No.~#1} 
\newcommand{\Seiten}[2]{\mbox{pp. #1--#2}} 

\newcommand{\Sueszmann}[0]{Sü{\normalfont ß}mann}
\newcommand{\Renyi}[0]{R\'enyi}

\newtheorem{Definition}{Definition} 
\newtheorem{Theorem}{Theorem}

\begin{document}
\title{Quantum phase space functions and relations of entropic localisation measures}
\author{Kedar S Ranade}\ead{Kedar.Ranade@uni-ulm.de}
\address{Institut f\"ur Quantenphysik, Universit\"at Ulm, and Center for Integrated Quantum Science and Technology
  (IQ\textsuperscript{ST}), Albert-Einstein-Allee 11, D-89081 Ulm, Deutschland (Germany)}
\date{May 31st, 2013}

\begin{abstract}
  The concept of quantum phase space offers a view on quantum mechanics, which is different from the
  standard Hilbert space approach, but which more closely resembles the classical phase space. Due
  to the properties of quantum mechanics there are several equivalent quantum phase space descriptions,
  and one cannot always prefer one or another as they all have certain merits and drawbacks. For example,
  the \pers{Husimi}-\pers{Kano} $Q$ function is a probability distribution and thus gives rise to
  entropic quantities, namely the \pers{\Renyi}-\pers{Wehrl} entropies, of which several properties
  are known. The \pers{Wigner} function, on the other hand, has an easier physical explanation, but
  may take negative values. In this article, we investigate entropic measures of localisation for a
  state in quantum phase space by using the \pers{Beckner}-\pers{Brascamp}-\pers{Lieb} inequality to
  relate different phase-space functions.
\end{abstract}

\maketitle

\section{Introduction}
In classical physics, the concept of a phase space is well-known and widely applied, e.\,g. in Lagrangian
and Hamiltonian mechanics. A (pure) state of such a system may be described by a point in phase space,
and the dynamics is given by a trajectory in this space. Statistical ensembles of classical states
may accordingly be described by a probability distribution on the phase space, and the dynamics of points
induces dynamics of such probabilities. On the other hand, the standard approach to quantum
mechanics uses the concept of a Hilbert space with vectors in and operators on the Hilbert space. But there also
exists the phase-space approach, which is arguably more similar to classical physics and exhibits different
phenomena of quantum mechanics. Mathematically, the quantum phase space formalism is an equivalent formulation
of quantum mechanics, which adapts classical phase space to quantum mechanics.
In this article we present a method to relate different entropic quantities on quantum phase space to each other
and to Hilbert space concepts such as purity by using tools from functional analysis.
\par The rest of this article is organised as follows: In section \ref{Sec_QuPhSp}, we introduce the basics of
quantum phase space distributions and discuss entropies and entropy-related measures and their interpretation.
In section \ref{Sec_Relations} we derive relations of these quantities and give some examples. We finally
conclude our discussion in section \ref{Sec_Conclusion} and offer an outlook on future work. In the appendix
we summarise concepts and results from quantum mechanics and mathematics, which are used in the text.

\section{Quantum phase space distributions and entropic measures of localisation}\label{Sec_QuPhSp}
Let us consider a single particle with position $x$ and momentum $p$, whose classical phase space is $\R \times \R$.
In the quantum case we take the same mathematical space, but have to fulfil e.\,g. the Heisenberg uncertainty relation,
so that a pure state cannot be a point (delta distribution). We can introduce quantum phase space in the following
way: we consider a state (density matrix) $\rho$ and combine $x$ and $p$ into $\alpha \in \C$ (cf. \ref{App_QM});
provided the expression is well-defined, we define the $s$-ordered phase space functions, $s \in \C$, by
(cf. e.\,g. \cite{Schleich} and also \cite{Puri} with different sign convention for $s$)
\begin{equation}\label{sFaltung}
  \PhV^{(s)}_\rho(\alpha) = \frac{1}{\pi^2} \int_{\ksi \in \C} \,\Spur\left(\rho\,\EZ^{\ksi\erz - \ksi^*\vnt}\right)\,
  \EZ^{-(\ksi\alpha^* - \ksi^*\alpha)}\,\EZ^{+\frac{s\betrag{\ksi}^2}{2}}\,\df{{}^2\ksi}.
\end{equation}
With $\D_s(\ksi) := \EZ^{\ksi\erz - \ksi^*\vnt +\frac{s\betrag{\ksi}^2}{2}}$, the \anfEngl{expectation value}
function $\ksi \mapsto \Spur \rho\,\D_s(\ksi)$ is known as the $s$-ordered characteristic function, and
$\PhV^{(s)}$ is its symplectic Fourier transform. The three most prominent phase-space functions
are the \pers{Glauber}-\pers{Sudarshan} $P$ function ($s = +1$), the \pers{Wigner} function $W$ ($s = 0$) and
the \pers{Husimi}-\pers{Kano} $Q$ function ($s = -1$). While $P$ need not even be a true function, $W$
and $Q$ are real-valued functions.\footnote{Another common definition of the \pers{Wigner} function is
$W_\rho(x,\,p) := \frac{1}{2\pi\hbar} \int_{\ksi \in \R} \bra{x + \frac{\ksi}{2}}\rho
\ket{x - \frac{\ksi}{2}}\EZ^{-\frac{\iE}{\hbar}p\ksi}\,\df{\ksi}$ \cite[p.~68]{Schleich}, which is easier to
interpret, but has the disadvantage that it has a dimension of inverse action; by relating phase-space domains
with $\dxdp = 2\hbar\da$, we find $W_\rho(\alpha) = 2\hbar \cdot W_\rho(x,\,p)$.}
\par As all phase space distributions contain the full information on the density matrix, they can be transformed
into each other. In particular, if $\Re(s) < \Re(t)$, there holds \cite[eq. (4.20)]{Puri}
\begin{equation}\label{PhasenRaumUmrechnung}
  \PhV^{(s)}(\alpha) = \frac{2}{\pi(t-s)} \int_{\beta \in \C} \PhV^{(t)}(\beta)
    \exp\left[-2\,\frac{\betrag{\alpha - \beta}^2}{t-s} \right] \,\df{^2\beta};
\end{equation}
this can be written as a two-dimensional convolution, $\PhV^{(s)} = \frac{2}{\pi(t-s)} \PhV^{(t)} * f_\frac{2}{t-s}$
with a Gaussian $\Fkt{f_a}{\C}{\R^+}$, $f_{a}(z) := \EZ^{-a\betrag{z}^2}$ and may be seen as a continuous averaging
or smoothing of $\PhV^{(t)}$.
\par We shall now introduce some measures of localisation of a quantum state, which give rise to entropic quantities.

\subsection{The \persbf{\Renyi}-Wehrl entropies}
The \pers{Husimi}-\pers{Kano} $Q$ distribution can be written in the form $Q_\rho(\alpha) = \frac{1}{\pi}
\bra{\alpha} \rho \ket{\alpha}$, which explicitly shows that it is non-negative and normalised, so that it can
directly be used to define entropies; for simplicity we shall always use the natural logarithm.
\begin{Definition}[\pers{\Renyi}-\pers{Wehrl} entropies \cite{GnutzmannZyczkowski}]\hfill\\
  For a density operator $\rho$ on $\cH = L^2(\R)$, the \pers{\Renyi}-\pers{Wehrl} entropy of order
  $q \in \R^+ \setminus \Mg{1}$ is defined~by
  \begin{align*}
    R_q(\rho) := \frac{1}{1-q} \ln \int_{\alpha \in \C} Q_\rho(\alpha)^q \da,
  \end{align*}
  and for $q \in \Mg{0,\,1,\,\infty}$ we take the appropriate limit. In particular, we get the (standard)
  \pers{Wehrl} entropy $W(\rho) := R_1(\rho)
  = -\int_{\alpha \in \C} Q_\rho(\alpha) \ln Q_\rho(\alpha) \da$.\footnote{Originally, the \pers{Wehrl} entropy 
  was defined by $\widetilde{W}(\rho) := -\frac{1}{\pi} \int_{\alpha \in \C} \widetilde{Q}_\rho(\alpha)
  \ln \widetilde{Q}_\rho(\alpha) \da = W(\rho) - \ln \pi$ using the non-normalised \pers{Husimi} function
  $\widetilde{Q}_\rho(\alpha) := \bra{\alpha}\rho\ket{\alpha}$ \cite{Wehrl,Lieb}. Note, however, that the limit
  $q \rightarrow 1$ in our definition is only sensible due to the normalisation of $Q_\rho$.}
\end{Definition}        
A \pers{\Renyi}-\pers{Wehrl} entropy is thus the \pers{\Renyi} entropy \cite{Renyi} of the $Q$ function or
the entropy with respect to the frame of coherent states (cf. \ref{App_QM}).
To illustrate this, we calculate \pers{\Renyi}-\pers{Wehrl} entropies for a general squeezed state
$\rho_{\alpha,\,\ksi} = \pr{\alpha,\,\ksi}$:
we have $R_q(\rho_{\alpha,\,\ksi}) = \frac{\ln q}{q-1} + \ln \pi + \ln \cosh \betrag{\ksi}$, and this splits into
three parts: (i) a \pers{\Renyi} part (with appropriate limits), (ii) a constant part $\ln \pi$ and (iii) a squeezing
part, which is related to the maximum absolute overlap
$\sqrt{\mathrm{sech} \betrag{\ksi}}$ of coherent states and states squeezed by $\Sq(\ksi)$.\footnote{Note that this
is an instance of the \pers{Wehrl}-\pers{Lieb} inequality \cite{Wehrl,Lieb,OhyaPetz}, which states that
$W(\rho) \geq 1 + \ln \pi$ with equality, if and only if $\rho$ is a coherent state.}

\subsection{The Süßmann measure}
Another measure of localisation of a quantum state in phase space is \pers{\Sueszmann}'s \emph{uncertainty area}
with dimension of an action, which is defined by\footnote{Alternatively, $\delta_\rho = \left(\int W_\rho(x,p)^2
\dxdp\right)^{-1}$ and $\Spur(\rho_1\rho_2) = 2\pi\hbar \int_{x,\,p \in \R} W_{\rho_2}(x,\,p) W_{\rho_2}(x,\,p) \dxdp$.}
\cite{Sueszmann,Schleich}
\begin{equation}
  \delta_\rho := \frac{2\hbar}{\int W_\rho(\alpha)^2 \da},
\end{equation}
from which the \pers{\Renyi}-\pers{\Sueszmann} entropy $S_\delta := \ln [(2\pi\hbar)^{-1} \delta_\rho]$ can be
constructed~\cite{Wlodarz}. Using $\Spur(\rho_1\rho_2) = \pi\int_{\alpha \in \C} W_{\rho_1}(\alpha)
W_{\rho_2}(\alpha) \da$ \cite[p.~71]{Schleich} we see that the \pers{\Sueszmann} measure is directly related
to the purity $\Spur(\rho^2)$ or the linear entropy $1 - \Spur(\rho^2)$ of the state: there holds
$\delta_\rho \geq 2\pi\hbar$ with equality, if and only if $\rho$ is pure.

\section{Relations of phase space functions}\label{Sec_Relations}
We shall now present a relation of phase-space quantities for different phase-space functions $\PhV^{(s)}$,
in particular the \pers{Wigner} and the \pers{Husimi} functions. Using the concept of $p$-norms
(cf. \ref{App_FunkAna}), we can rewrite the \pers{\Renyi}-\pers{Wehrl} entropies as
\begin{align}
  R_q(\rho) = \frac{q}{1-q} \ln \norm{Q_\rho}_q, \quad q > 1,
\end{align}
and the \pers{\Renyi}-\pers{\Sueszmann} entropy as
\begin{align}
  S_{\delta_\rho} = \ln \left(\pi \int W_\rho(\alpha)^2 \da\right)^{-1}
    = \ln \left(\frac{1}{\pi} \norm{W_\rho}_2^{-2}\right) = - \ln \pi - 2 \ln  \norm{W_\rho}_2.
\end{align}
Solving for the norm of the phase-space function these equations read
\begin{align}
  \norm{Q_\rho}_q = \exp\left(\frac{1-q}{q} R_q(\rho)\right) \quad\text{and}\quad
  \norm{W_\rho}_2 = \exp\left[-\frac{1}{2}(S_{\delta_\rho} + \ln \pi)\right].
\end{align}
Now, we can use $Q = \frac{2}{\pi}(W * f_2)$ and plug this into the \pers{Beckner}-\pers{Brascamp}-\pers{Lieb}
inequality (Theorem~\ref{BBL}) for $p = 2$. Then, for $1 + \frac{1}{r} = \frac{1}{2} + \frac{1}{q}$ or
$q = \frac{2r}{r+2}$, it follows
\begin{align}\label{EntropyRelation}
  \EZ^{\frac{1-r}{r} R_r(\rho)} = \norm{Q_\rho}_r
  &\leq \left(\frac{C_q}{C_r}\right)^2 \cdot \frac{2}{\pi} \cdot \norm{W_\rho}_2 \norm{f_2}_q\\
  &= \left(\frac{C_q}{C_r}\right)^2 \cdot \frac{2}{\pi} \cdot
    \EZ^{-\frac{1}{2}(S_{\delta_\rho} + \ln \pi)} \cdot \left(\frac{\pi}{2q}\right)^\frac{1}{q}
\end{align}
We can now consider special cases of this inequality and start by mentioning that $r \geq 2$ is necessary for
$q \geq 1$. For the case of $r = 2$, the \pers{\Renyi}-\pers{Wehrl} collision entropy, we find $q = 1$, and the
expression reduces to
\begin{equation}
  \EZ^{-\frac{1}{2} R_2(\rho)} \leq \EZ^{-\frac{1}{2}(S_{\delta_\rho} + \ln \pi)},
  \quad\text{i.\,e.}\quad S_{\delta_\rho} + \ln \pi \leq R_2(\rho).
\end{equation}
The other main example is $r \rightarrow \infty$, from which there follows $q = 2$, so that eq.~(\ref{EntropyRelation})
then reads
\begin{equation}
  \EZ^{-R_\infty(\rho)} \leq \pi^{-1/2} \cdot \EZ^{-\frac{1}{2}(S_{\delta_\rho} + \ln \pi)},
  \quad\text{i.\,e.}\quad 2 R_\infty(\rho) \geq S_{\delta_\rho} + 2\ln \pi.
\end{equation}
In this case, we can explicitly see that equality holds for coherent states.

\subsection{Generalisations}
We do not need to restrict to the case of the \pers{\Sueszmann} measure only, but can consider all quantities
$\norm{W_\rho}_p$ for $p \in [1;\,\infty]$. Note that none of these quantities change under Gaussian
operations such as displacement and squeezing or with time in free evolution or in a harmonic potential.
\par For example, in the case that $p = 1$, the normalisation of the \pers{Wigner} function,
$\int_{\alpha \in \C} W_\rho(\alpha) \da = 1$, implies $\norm{W_\rho}_1 \geq 1$ with equality, if and only
if the \pers{Wigner} function is non-negative. As the \pers{Wigner} function is often considered to be a
\anfEngl{non-classical} probability distribution, we may e.\,g.
define $C_\rho := \ln \norm{W_\rho}_1$ as the \anfEngl{non-classicality} of a quantum state.\footnote{As an
example, there roughly holds $\norm{W_{\pr{m}}}_1 \approx \sqrt{m+1}$ for \pers{Fock} states, so that
$C_\rho \approx \frac{m+1}{2}$.} With $p = 1$ and thus $r = q$, our inequality reads
\begin{equation}
  \norm{Q_\rho}_q \leq \frac{2}{\pi} \cdot \norm{W_\rho}_1 \cdot \norm{f_2}_q
  \quad\text{with}\quad \norm{f_2}_q = \left(\frac{\pi}{2q} \right)^{1/q}.
\end{equation}
In the case of $q \in \Mg{1,\,2,\,\infty}$ the the right-hand side factor becomes $\pi/2$, $\sqrt{\pi}/2$ and $1$, respectively.
\par Another example is $p = \infty$, which yields the maximum of the \pers{Wigner} function on phase space. In this
case, we have $1 + \frac{1}{r} = \frac{1}{q}$ or $q = \frac{r}{r+1}$, so that only $r = \infty$ and $q = 1$ remains
possible, and the inequality then reads $\norm{Q_\rho}_\infty \leq \norm{W_\rho}_\infty$.

\subsection{Perspectives}
It might be possible to prove the \pers{Hudson}-\pers{Piquet} theorem \cite{Schleich}, which states that a pure
state has a non-negative \pers{Wigner} function, if and only if it is Gaussian, in a functional-analytic way
(the standard proofs use methods from the theory of functions \cite{Hudson,Piquet}):
By construction, both the \pers{Wigner} and the \pers{Husimi} function are normalised in the sense
that $\int_{\alpha \in \C} W(\alpha) \da = \int_{\alpha \in \C} Q(\alpha) \da = 1$. As the latter one
is non-negative everywhere, there also holds $\norm{Q}_1 = 1$, but this is not true for the former. 
There holds $\norm{W}_1 \geq 1$ with equality, if and only if $W$ is non-negative, i.\,e. $W \geq 0$.
Using $Q = \frac{2}{\pi}(W * f_2)$, from Theorem \ref{BBL} there follows $1 = \norm{Q}_1 \leq \frac{2}{\pi} \norm{W}_1
\norm{f_2}_1 = \norm{W}_1$, and we have to check the conditions for equality. The purity of the state
can be reformulated in norms by $\norm{W_\rho}_2 = \pi^{-1/2}$. However, the conditions of equality from Theorem \ref{BBL}
do not apply here.

\section{Conclusion}\label{Sec_Conclusion}
In this work, we have found relations between entropic quantities in quantum phase space, in particular, for
\pers{Wigner} and \pers{Husimi} functions, by using the \pers{Beckner}-\pers{Brascamp}-\pers{Lieb} inequality,
which points out specific properties of Gaussian functions in a natural way. Some of these
quantities, e.\,g. the \pers{\Sueszmann} measure, are directly related to Hilbert-space properties.
The relations which we found for \pers{Wigner} and \pers{Husimi} distributions can, in principle, be generalised
to other $s$-ordered phase-space functions, and one may hope to find more relations for different distributions.
\ack The author thanks \pers{Wolfgang Schleich} for helpful information on quantum phase space and
acknowledges funding by BMBF/QuOReP (Förderkennzeichen 01BQ1014).

\appendix

\section{Quantum mechanics: Harmonic oscillators and Gaussians}\label{App_QM}
A single spinless, pointlike particle in a one-dimensional harmonic oscillator is described by the separable,
infinite-dimensional complex system Hilbert space \mbox{$\cH = L^2(\R)$}. The \emph{position} and \emph{momentum
operators}, $\hat{x}$ and $\hat{p}$, fulfil the  \emph{canonical commutation relation}
$[\hat{x},\,\hat{p}] = \iE\hbar\Eins_\cH$; in position representation, they read $\hat{x} = x$ and
$\hat{p} = -\iE\hbar\nabla$, and the Hamiltonian operator is $\hat{H} = \frac{\hat{p}^2}{2m} + \hat{V}
= -\frac{\hbar^2}{2m} \Delta + V(x)$ with potential $V(x) := \frac{1}{2}m\omega x^2$.
\par Classical position and momentum are combined into a dimensionless parameter
$\alpha = \sqrt{\frac{m\omega}{2\hbar}}\,x + \frac{\iE}{\sqrt{2m\omega\hbar}}\,p \in \C$ with differentials
$\df{x}\,\df{p} = 2\hbar\,\df{\Re(\alpha})\,\df{\Im(\alpha}) = 2\hbar\,\df{}^2 \alpha$.
By \anfEngl{canonical quantisation}---replacing $(x,\,p)$ by $(\hat{x},\,\hat{p})$---$\alpha$ turns
into an operator~$\vnt$ with $[\hat{x},\,\hat{p}] = \iE\hbar \eqv [\vnt,\erz] = 1$.
On $\cH$ with the orthonormal \pers{Fock} (or photon-number) basis $\Mge{\ket{n}}{n \in \N_0}$,
$\erz$ and $\vnt$ are the unbounded \emph{creation} and \emph{annihilation operators} with spectrum $\emptyset$
and $\C$, respectively, where $\erz \ket{n} = \sqrt{n+1} \ket{n+1}$ and $\vnt \ket{n} = \sqrt{n} \ket{n-1}$, except
for $\vnt \ket{0} = 0$. The \emph{number operator} $\anz = \erz\vnt$ with $\anz\ket{n} = n \ket{n}$ has spectrum
$\N_0$, and $\hat{H} = \hbar\omega(\anz + \frac{1}{2}) = \hbar\omega(\erz\vnt + \frac{1}{2})$,
related to the classical overall energy $E_\mathrm{tot} = \hbar\omega\alpha^*\alpha$.
\par Two systems of unitary operators, the \emph{displacement operators} $\D(\alpha) := \EZ^{\alpha\erz-\alpha^*\vnt}$
for $\alpha\in \C$ and the \emph{squeezing operators} $\Sq(\ksi) := \EZ^{-\frac{1}{2}(\ksi {\erz}{}^2 - \ksi^* \vnt^2)}$
for $\ksi = r \EZ^{\iE\phi} \in \C$, define \emph{coherent states} $\ket{\alpha} := \D(\alpha)\ket{0}$
(note that $\vnt\ket{\alpha} = \alpha\ket{\alpha}$) and
\emph{squeezed vacuum states} $\ket{\ksi} := \Sq(\ksi)\ket{0}$, respectively. The set $\Mge{\ket{\alpha}}{\alpha \in \C}$
does not form an orthonormal basis, but somewhat similar, by $\int_{\alpha \in \C} \ket{\alpha} \bra{\alpha} \da = \pi\Eins$
(pointwise), an overcomplete tight continuous \emph{frame} with frame bound $\pi$~\cite{Christensen} and,
more generally, a \emph{positive operator-valued measure} (POVM) \cite{Berberian,Ruskai}.
The \emph{two-mode} and \emph{ideal} \emph{squeezed (coherent) states} $\ket{\alpha,\,\ksi}$ are defined by
$\Sq(\ksi)\D(\alpha)\ket{0}$ or $\D(\alpha)\Sq(\ksi)\ket{0}$, respectively \cite[p.~1042]{MandelWolf}, and are related by
$\Sq(\ksi)\D(\alpha) = \D(\beta)\Sq(\ksi)$ with
$\beta = \alpha \cosh r + \alpha^* \EZ^{\iE\phi} \sinh r$ \cite[p. 18 with different signs]{DrummondFicek}.
Due to $\D(\alpha + \beta) = \D(\alpha) \D(\beta)\,\EZ^{\frac{\alpha^*\beta-\alpha\beta^*}{2}}
= \D(\alpha) \D(\beta)\,\EZ^{\iE(\Re\,\alpha\,\Im\,\beta-\Im\,\alpha\,\Re\,\beta)}$ the map
$\alpha \mapsto \D(\alpha)$ is a unitary projective representation of the abelian group $(\C,+) \cong (\R^2,+)$;
although $\Sq(\ksi)^{-1} = \Sq(-\ksi)$, this is not true for the squeezing operators due to
$[\vnt^2,\,\vnt^{\dagger 2}] = 2 \{\vnt,\,\erz\} = 4(\anz + \frac{1}{2}) \neq \mathrm{const.}$

\section{Functional analysis: Norms, convolutions and some integrals}\label{App_FunkAna}
In order to derive relations between entropic measures, we use some well-known mathematical concepts from
functional analysis, which we will review here. The most important concept in our calculations is the $p$-norms
(cf. e.\,g. \cite{Werner,Conway}).
\begin{Definition}[$p$-norms]\hfill\\
  For a measure space $(X,\my)$---in particular, for $X = \R^n$ with the Lebesgue measure---and a measurable
  function $\Fkt{f}{X}{\C}$, the $p$-norm $\Fkt{\norm{\,\cdot\,}_p}{X}{\R^+_0}$ is (provided the expression
  is finite) defined by $\norm{f}_p := \left(\int_{x \in X} \betrag{f(x)}^p \,\df{\my(x)}\right)^{1/p}$, if
  $p \in [1;\infty)$, and $\norm{f}_\infty := \mathrm{ess} \sup_{x \in X} \betrag{f(x)} := \inf\Mge{a \in \R}
  {\my(\Mge{x \in X}{\betrag{f(x)} > a}) = 0}$.
\end{Definition}
Note that $\norm{f}_p = \norm{g}_p \cdot \norm{h}_p$ for $f(x,\,y) = g(x) \cdot h(y)$.
The convolution of functions $\Fkt{f,\,g}{\R^n}{\C}$ is given by $(f * g)(x) := \int_{y \in \R^n}
f(x-y)g(y)\,\df{y}$. For a one-dimensional Gaussian function $f_a(x) := \EZ^{-ax^2}$ with $a > 0$ we find
$(f_a * f_b)(x) := \sqrt{\frac{\pi}{a+b}} f_{\frac{ab}{a+b}}$, i.\,e., for a Gaussian normal distribution
$N_{\my,\,\sigma^2}(x) := \frac{1}{\sigma\sqrt{2\pi}} \exp [-\frac{1}{2}(\frac{x-\my}{\sigma})^2]$ with mean
value $\my \in \R$ and variance $\sigma^2 \in \R^+$ (standard deviation $\sigma := \sqrt{\sigma^2}$):
$N_{\my_1,\,\sigma_1^2} * N_{\my_2,\,\sigma_2^2} = N_{\my_1 + \my_2,\,\sigma_1^2 + \sigma_2^2}$.
\par We can now discuss a strengthened \pers{Young}'s inequality (\cite[p. 169]{Beckner},
\cite[p. 168f.]{BrascampLieb}): The conjugate of $p \in [1;\infty]$ is $p^\prime = \frac{p}{p-1}$
(with $\frac{1}{0} := \infty$ and vice versa), so that $\frac{1}{p} + \frac{1}{p^\prime} = 1$ and
$1 + \frac{1}{r} = \frac{1}{p} + \frac{1}{q}$ is equivalent to $\frac{1}{r^\prime} = \frac{1}{p^\prime}
+ \frac{1}{q^\prime}$. We set $C_p := \sqrt{p^{1/p} / {p^\prime{}^{1/p^\prime}}}$ with $C_1 := C_\infty := 1$;
note that $C_p C_{p^\prime} = 1$ and, in particular, $C_2 = 1$.
\begin{Theorem}[Beckner-Brascamp-Lieb inequality]\label{BBL}\hfill\\
  For $p,\,q,\,r \in [1;\infty]$ such that $1 + \frac{1}{r} = \frac{1}{p} + \frac{1}{q}$ and functions
  $\Fkt{f,\,g}{\R^n}{\C}$, there holds $\norm{f * g}_r \leq (C_p C_q C_r^{-1})^n \norm{f}_p \norm{g}_q$,
  with equality, for $n = 1$ and $p,\,q \in (1;\infty)$, if and only if
  $f(x) = A\,\EZ^{-\gamma p^\prime(x-\alpha)^2+\iE\delta x}$ and
  $g(x) = B\,\EZ^{-\gamma q^\prime(x-\beta)^2+\iE\delta x}$
  for some $A,\,B \in \C$, $\alpha,\,\beta,\,\delta \in \R$ and $\gamma \in \R^+$.
\end{Theorem}
We finally remind the reader that all integrals over $\C$ correspond to integrals over~$\R^2$. For
$a,\,b,\,c \in \C$ with $\Re(a) > 0$, there holds $\int_{x \in \R} \EZ^{-ax^2+bx+c}\,\df{x}
= \sqrt{\frac{\pi}{a}}\,\EZ^{\frac{b^2}{4a} + c}$, where the square root takes positive real part,
and $\int_{x = 0}^{\infty} x^{m-1} \EZ^{-ax^2} \,\df{x} = \frac{\Gamma(\frac{m}{2})}{ 2a^{\frac{m}{2}}}$
for $m \in \C$ with $\Re(m) > 0$.
In particular, for $f_a(x) := \EZ^{-ax^2}$, there holds $\norm{f_a}_p = (\frac{\pi}{pa})^{n/2p}$.

\begin{flushleft}

\end{flushleft}

\end{document}